\documentclass[sigconf]{acmart}
\AtBeginDocument{%
  \providecommand\BibTeX{{%
    \normalfont B\kern-0.5em{\scshape i\kern-0.25em b}\kern-0.8em\TeX}}}

\copyrightyear{2024}
\acmYear{2024}
\setcopyright{acmlicensed}\acmConference[CIKM '24]{Proceedings of the 33rd ACM International Conference on Information and Knowledge Management}{October 21--25, 2024}{Boise, ID, USA}
\acmBooktitle{Proceedings of the 33rd ACM International Conference on Information and Knowledge Management (CIKM '24), October 21--25, 2024, Boise, ID, USA}
\acmDOI{10.1145/3627673.3680015}
\acmISBN{979-8-4007-0436-9/24/10}

\usepackage{algorithm}
\usepackage{algorithmic}
\usepackage{pdfpages}
\usepackage{multirow}
\usepackage[export]{adjustbox} 
\usepackage{bm}
\usepackage{amsmath}
\usepackage{array}
\usepackage{booktabs}
\usepackage{footmisc}

\usepackage{amsthm}
\usepackage{graphicx}
\usepackage{enumitem}

\theoremstyle{definition}

\usepackage{subfigure}
\usepackage{hyperref}
\usepackage{url}
\usepackage{color}
\usepackage{xcolor}
\usepackage{colortbl}
\usepackage{xspace}

\newcommand{\model}{FRGCF\xspace}

\usepackage{amsmath,amsfonts,bm}

\def\eqref#1{equation~\ref{#1}}
\def\1{\bm{1}}

\def\mE{{\bm{E}}}

\DeclareMathAlphabet{\mathsfit}{\encodingdefault}{\sfdefault}{m}{sl}
\SetMathAlphabet{\mathsfit}{bold}{\encodingdefault}{\sfdefault}{bx}{n}

\def\gG{{\mathcal{G}}}

\def\gI{{\mathcal{I}}}

\def\gL{{\mathcal{L}}}

\def\gU{{\mathcal{U}}}

\begin{document}

\title{Feedback Reciprocal Graph Collaborative Filtering}


\author{Weijun Chen}
\authornote{Both authors contributed equally to this research.}
\affiliation{%
 \institution{Jinan University}
 %\streetaddress{Rono-Hills}
 \city{Guangzhou}
 %\state{Arunachal Pradesh}
 \country{China}}
\email{chenweijunjnu@gmail.com}

\author{Yuanchen Bei}
\authornotemark[1]
\affiliation{%
 \institution{Zhejiang University}
 %\streetaddress{Rono-Hills}
 \city{Hangzhou}
 %\state{Arunachal Pradesh}
 \country{China}}
\email{yuanchenbei@zju.edu.cn}

\author{Qijie Shen}
\affiliation{%
 \institution{Alibaba Group}
 %\streetaddress{Rono-Hills}
 \city{Hangzhou}
 %\state{Arunachal Pradesh}
 \country{China}}
\email{qijie.sqj@alibaba-inc.com}

\author{Hao Chen}
\authornote{Corresponding author.}
\affiliation{%
 \institution{The Hong Kong Polytechnic University}
 %\streetaddress{Rono-Hills}
 \city{Hung Hom}
 %\state{Arunachal Pradesh}
 \country{Hong Kong SAR}}
\email{sundaychenhao@gmail.com}

\author{Xiao Huang}
\affiliation{%
 \institution{The Hong Kong Polytechnic University}
 %\streetaddress{Rono-Hills}
 \city{Hung Hom}
 %\state{Arunachal Pradesh}
 \country{Hong Kong SAR}}
\email{xiaohuang@comp.polyu.edu.hk}

\author{Feiran Huang}
\affiliation{%
  \institution{Jinan University}
  \city{Guangzhou}
  \country{China}
}
\email{huangfr@jnu.edu.cn}

\renewcommand{\shortauthors}{Weijun Chen, et al.}

\begin{abstract}
Collaborative filtering on user-item interaction graphs has achieved success in the industrial recommendation. 
However, recommending users' truly fascinated items poses a seesaw dilemma for collaborative filtering models learned from the interaction graph.
On the one hand, not all items that users interact with are equally appealing. Some items are genuinely fascinating to users, while others are unfascinated. Training graph collaborative filtering models in the absence of distinction between them can lead to the recommendation of unfascinating items to users.
On the other hand, disregarding the interacted but unfascinating items during graph collaborative filtering will result in an incomplete representation of users' interaction intent, leading to a decline in the model's recommendation capabilities. To address this seesaw problem, we propose \underline{F}eedback \underline{R}eciprocal \underline{G}raph \underline{C}ollaborative \underline{F}iltering (\model), which emphasizes the recommendation of fascinating items while attenuating the recommendation of unfascinating items.
Specifically, \model first partitions the entire interaction graph into the Interacted \& Fascinated (I\&F) graph and the Interacted \& Unfascinated (I\&U) graph based on the user feedback.
Then, \model introduces separate collaborative filtering on the I\&F graph and the I\&U graph with feedback-reciprocal contrastive learning and macro-level feedback modeling. This enables the I\&F graph recommender to learn multi-grained interaction characteristics from the I\&U graph without being misdirected by it.
Extensive experiments on four benchmark datasets and a billion-scale industrial dataset demonstrate that \model improves the performance by recommending more fascinating items and fewer unfascinating items.
Besides, online A/B tests on Taobao's recommender system verify the superiority of \model.
\end{abstract}





\maketitle

\section{Introduction}
Motivated by the success of graph neural networks (GNNs) in tasks on the relational data~\cite{kipf2016semi,hamilton2017inductive,wu2022graph,shengyuan2024differentiable,bei2024cpdg}, graph-based collaborative filtering (GCF) has formulated user-item historical interactions as user-item graphs and then recursively aggregated interacted neighbors' representations to obtain the refined embeddings of each user and item~\cite{he2020lightgcn,wu2021self,zhou2023adaptive}. 

\begin{figure}[tbp]
    \centering
    \includegraphics[width=\linewidth, trim=0cm 0cm 0cm 0cm,clip]{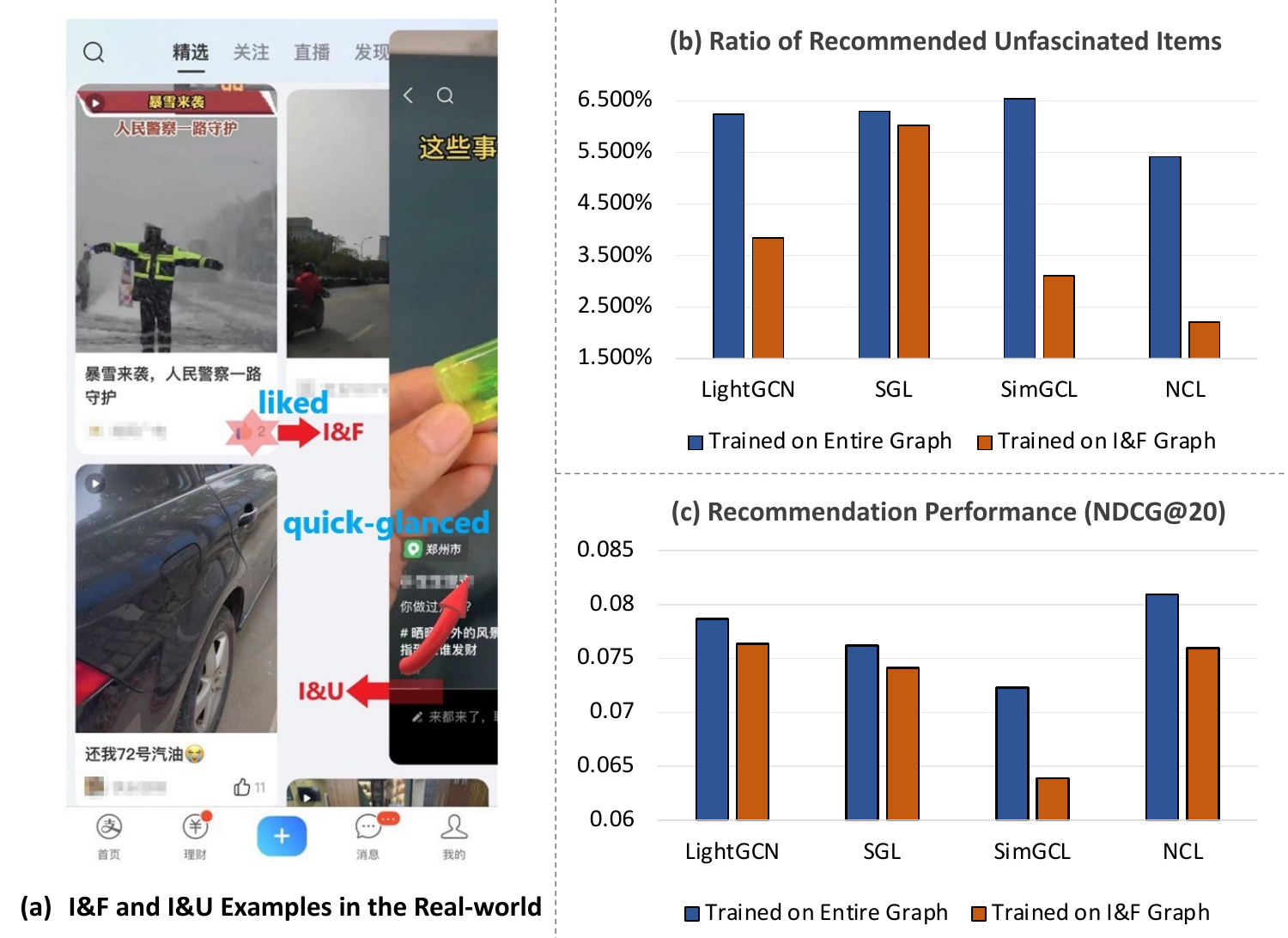}
    \caption{(a) The Interacted \& Fascinated (I\&F) and Interacted \& Unfascinated (I\&U) examples. (b) The recommended unfascinated items of GCF models between trained on the entire graph and the I\&F graph. (c) The recommendation performance difference of GCF models between trained on the entire graph and the I\&F graph.}
    \vspace{-0.55em}
    \label{fig:intro}
\end{figure}

However, learning meaningful representations from the interaction graph to recommend items that truly fascinate users remains a seesaw dilemma.
On the one hand, \textbf{not all interacted items are attractive to users}.
For example, as shown in Figure \ref{fig:intro}-(a), the recommended short videos with exquisite content are more likely to fascinate users and \textit{receive higher ratings or likes feedback}~\cite{xie2021deep,zhang2023contrastive} (short as \textbf{I\&F items}), while those videos that users are not actually interested in but with appealing covers and tiresome content may also be clicked by them but \textit{receive low rating scores or quick-glanced}~\cite{ajesh2016random,margaris2020makes} (short as \textbf{I\&U items}). 
Existing studies mainly directly train on the entire interaction graph, which does not consider and distinguish the difference between fascinated and unfascinated interactions.
In this manner, GCF models aggregate all interacted neighbors' information to represent users or items, where the information of unfascinated neighbors can also be aggregated. Then, during the recommendation process, the aggregated user embeddings have close relationships with the unfascinated items, and thus the users will be recommended a certain number of unfascinating items, leading to a negative user experience. As shown in Figure \ref{fig:intro}-(b), the recommendation results of GCF models trained on the entire interaction graph contain a high ratio of user-unfascinated items. 
On the other hand, \textbf{disregarding the I\&U interactions leads to an incomplete understanding of user-item interactions}.
Although the unfascinated items may bore users, they also reflect some intent of users, and directly ignoring all these items will lose potentially helpful information and their positive attributes that attract user interaction.
As shown in Figure \ref{fig:intro}-(c), the GCF models only trained on the I\&F graph have performance drops compared with the performance of models trained on the entire graph.
Therefore, it is vital to comprehensively consider users' interaction intent from both interactions with fascinated and unfascinated feedback while recommending users with more fascinated items and fewer unfascinated items.

Although important, building GCF models of this target presents two main challenges: 
(1) \textbf{Joint modeling of I\&F and I\&U interactions}: 
Both I\&F and I\&U interactions contain valuable information for user modeling. Simply separating interactions according to fascinated and unfascinated feedback to build separate models would greatly reduce the training data and negatively impact the performance~\cite{tang2020progressive,ma2022crosscbr}. Thus, it remains a challenge to comprehensively consider both two views of interactions.
(2) \textbf{Building connections between I\&F and I\&U interactions}:
Besides the joint modeling, there is a further challenge to establish an effective connection between the I\&F perspective and the I\&U perspective without being affected by the negative factors of I\&U interactions.

To address these issues, we propose a novel \underline{F}eedback \underline{R}eciprocal \underline{G}raph \underline{C}ollaborative \underline{F}iltering~(\textbf{\model}). Specifically, we first utilize two separate \textit{feedback-partitioned graph collaborative filtering} modules for the partitioned I\&F graph and I\&U graph. Further, we design a \textit{feedback-reciprocal contrastive learning} to connect the I\&F graph and I\&U graph, enabling better modeling of the user's interaction intent through the exchange of information while avoiding the model being misled by the I\&U interaction. Then, we design a \textit{macro-level feedback modeling} scheme to alleviate the data incompleteness caused by separately modeling I\&F graph and I\&U graph.
Finally, the recommender constructs on the I\&F graph can capture the interaction character of the I\&U graph without aggregating the information of the I\&U items, and thus is adopted for inference.
The main contributions are listed as follows:
\begin{itemize}[leftmargin=*]
    \item We highlight the seesaw dilemma in current graph collaborative filtering methods, which is largely neglected by previous works and poses difficulties in taking into account both the complete modeling of user intent and recommending more items that users are truly fascinated by.
    \item We propose \model, a novel approach using two different-intent subgraphs: I\&F graph and I\&U graph. \model separately models the I\&F graph and the I\&U graph with feedback-reciprocal contrastive learning and macro-level feedback modeling for meaningful feedback information learning and exchanging while avoiding the negative impacts of I\&U interactions. 
    \item Extensive experiments on four benchmark datasets and a billion-scale industrial dataset demonstrate that \model can outperform existing models while reducing the recommendation of unfascinating items. Online A/B studies verify the effectiveness of \model in Taobao's industrial recommender system.
\end{itemize}

\section{Related Works}

\subsection{Graph Collaborative Filtering}
Recently, the success of GNNs in modeling complex relational data has attracted widespread attention in various applications~\cite{kipf2016semi,hamilton2017inductive,zhang2024logical,bei2023reinforcement,zhang2023integrating}. Given that user-item interactions can inherently be perceived as a graph structure, numerous studies have focused on graph collaborative filtering (GCF) and achieved good recommendation performance~\cite{wang2019neural,he2020lightgcn,wu2022survey,zhang2024linear}.
Further, in recent endeavors to address the intricacies of noise and scarcity within GCF, a variety of self-supervised Learning (SSL) techniques have been explored and obtained state-of-the-art performance~\cite{yang2021enhanced,xia2021self,lin2022improving}. 
Current SSL-based GCF models can be mainly organized into two main paradigms: contrastive-based GCF and predictive-based GCF, all of which mainly share the common trait of employing augmented data views~\cite{wu2021self,xia2021self,lee2021bootstrapping,yu2022graph,lin2022improving}. The defined views can take the form of two embedding views in contrastive GCF or data views utilized for estimating mutual information in predictive GCF.

However, the dilemma of these GCF models remains, with some unfascinating items being recommended to users due to the training manner, with the entire interaction graph undistinguishing between the fascinating and unfascinating items. This will lead to the recommendation of unfascinating items to users and affect the user experience in real-world applications. In this paper, our designed \model is an SSL-based GCF model to tackle the highlighted seesaw dilemma with cooperative modeling of the I\&F and I\&U feedbacked interactions.

\subsection{User Feedback Enhanced Recommendation}
User feedback is an important signal for recommendation evaluation~\cite{chen2022generative,huang2023aligning,zhang2024multi,zhao2024recommender,chen2022differential}. Positive feedback from users indicates satisfaction with the item being interacted with, reflecting the users' genuine interest. While negative feedback represents dissatisfaction with the product, and continuing to recommend such products may affect the user experience.
In recent years, the explicit consideration of user feedback has received more and more attention in sequential recommendations and click-through rate prediction~\cite{xie2021deep,gong2022positive,fan2022modeling,chen2021curriculum,pan2023learning}. When modeling historical sequences, these models consider different weights for items with different user feedback. For GCF models, in recent years, a small amount of work has started to design customized models to take into account user feedback~\cite{seo2022siren,huang2023negative,hennekes2023weighted}. They incorporate both positive and negative encoders for different user feedback considerations. 

However, the seesaw dilemma in GCF of using both I\&F and I\&U graphs simultaneously or using only the I\&F graph and ignoring the I\&U graph has not yet been solved. Therefore, we design \model for addressing the seesaw problem in this paper.

\section{Methodology}
\textbf{Notations.}
We suppose the sets of users and items to be $\gU=\{u_{1}, u_{2}, ..., u_{M}\}$, and $\gI=\{i_{1}, i_{2}, ..., i_{N}\}$, where $M$ and $N$ are the number of users and items. The user-item interaction graph is defined as $\mathcal{G} = \{\mathcal{U}, \mathcal{I}, \mathcal{E}\}$, where $\mathcal{U}$ and $\mathcal{I}$ is the set of users and items, $\mathcal{E}$ is a set of interactions with user feedback.
Each interaction $r_{u, i} \in \mathcal{E}$ is denoted as a triple $(u, i, p)$, where $u \in \mathcal{U}$, $i \in \mathcal{I}$. $p$ describes the feedback of the user, such as the rating score, completion rate, and page stay time.

\begin{figure*}[tbp]
    \centering
    \includegraphics[width=\linewidth, trim=0cm 0cm 0cm 0cm,clip]{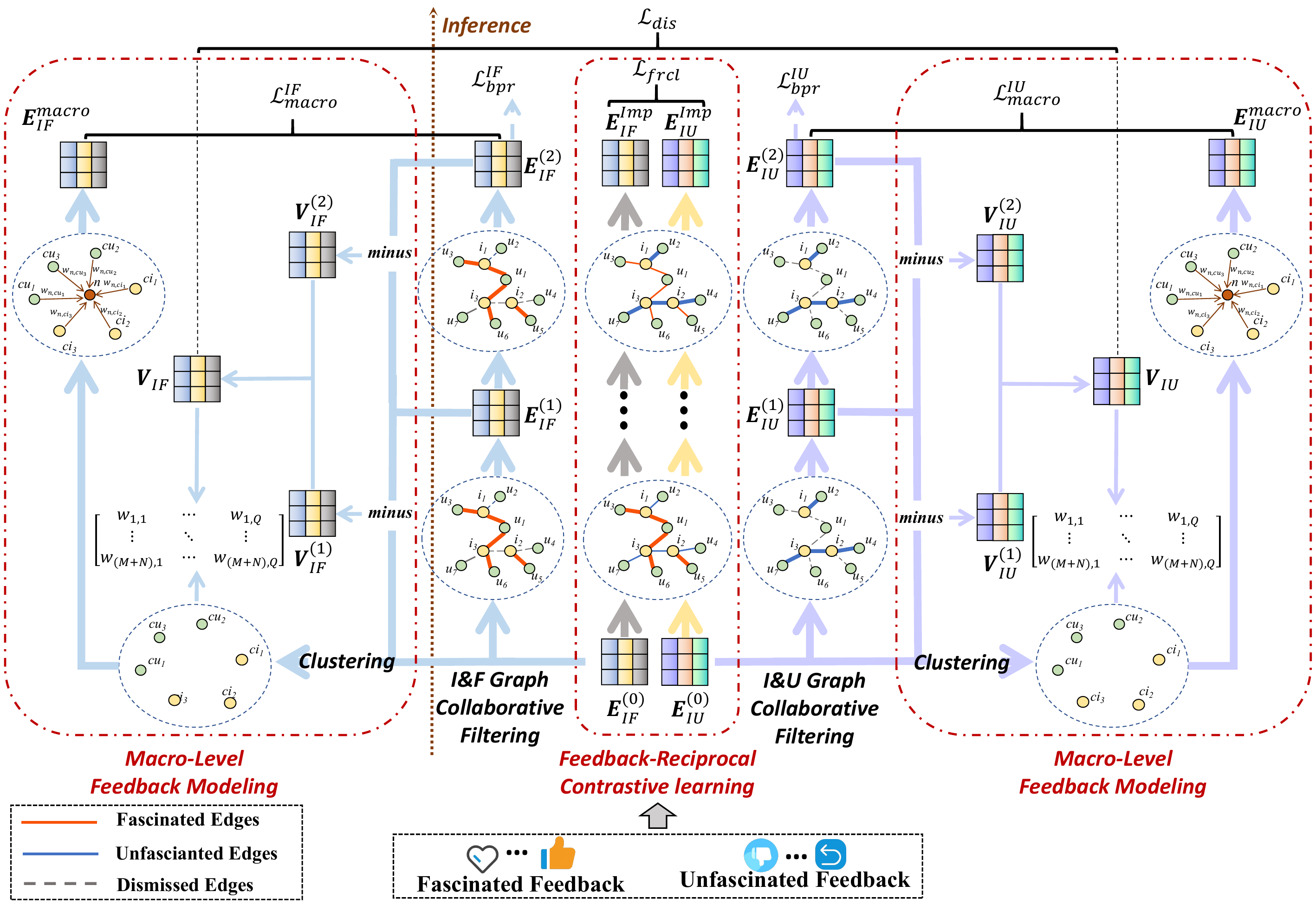}
    \caption{The overall framework of our \model, which contains three main components: (i) Feedback-partitioned graph collaborative filtering to train GCF models for I\&F graph and I\&U graph, respectively, and (ii) Feedback-reciprocal contrastive learning to connect the I\&F GCF model and the I\&U GCF model.
    (iii) Macro-level feedback modeling to enhance the representation of I\&F and I\&U. (iv) Finally, the inference predictions are computed by the I\&F graph convolution model.}
    \label{fig:framework}
\end{figure*}

As shown in~\autoref{fig:framework}, we first introduce the feedback-partitioned graph collaborative filtering for training the interaction prediction model with I\&F and I\&U GCF, respectively. Then we introduce the feedback-reciprocal contrastive learning to interchange the interaction information between the I\&F and I\&U graph collaborative filtering process. After that, we introduce the macro-level feedback modeling and distance regularization to enhance representation. Finally, we give the optimization and inference process of \model.

\subsection{Feedback-Partitioned Graph Collaborative Filtering}
In this subsection, we first describe the splitting of the whole interacted graph towards the Interacted \& Fascinated (I\&F) graph and the Interacted \& Unfascinated (I\&U) graph. Then we introduce the graph convolution for I\&F and I\&U graphs.

\subsubsection{\textbf{Interaction Graph Partition}}
The whole interaction graph $\gG$ is partitioned into the I\&F graph $\gG_{IF}$ and the I\&U graph $\gG_{IU}$ according to the user feedback towards items, which can be formally represented as:
\begin{equation}
    \gG_{IF} = \{\mathcal{U}, \mathcal{I}, \mathcal{E}_{IF}\},\quad \gG_{IU} = \{\mathcal{U}, \mathcal{I}, \mathcal{E}_{IU}\},
\end{equation}
where $\mathcal{E}_{IF} = \{(u, i, p) | p \in F^{pos} \}$ and $\mathcal{E}_{IU} = \{(u, i, p) | p \in F^{neg}\}$, $F^{pos}$ and $F^{neg}$ are the fascinated feedback set and unfascinated feedback set, respectively.

For the partitioned I\&F graph $\gG_{IF}$, we can construct the interaction matrix $\bm{R}_{IF}\in \mathbb{R}^{M \times N}$ from the edge set $\mathcal{E}_{IF}$, and then obtain the adjacency matrix $\bm{A}_{IF}$ of $\gG_{IF}$ as:
\begin{equation}
\bm{A}_{IF} = 
\begin{pmatrix}
\bm{0} & \bm{R}_{IF} \\
\bm{R}_{IF}^{T} & \bm{0}  \\
\end{pmatrix},\quad
\bm{A}_{IU} = 
\begin{pmatrix}
\bm{0} & \bm{R}_{IU} \\
\bm{R}_{IU}^{T} & \bm{0}  \\
\end{pmatrix}.
\end{equation}

\subsubsection{\textbf{I\&F/I\&U Graph Collaborative Filtering}}
The challenge of entire graph training lies in attempting to utilize a set of embeddings to accurately represent interactive behaviors that have multiple intricate interpretations. This is further supported by the findings depicted in Figure \ref{fig:intro}-(b). However, by simplifying the modeling focus of embeddings, higher-quality representations can be achieved. Consequently, we opted to establish two distinct sets of embeddings in order to model high-quality I\&F representations and I\&U representations respectively.

Regarding the dissemination of information within the $\gG_{IF}$, we can obtain the recursive graph collaborative layers by multiplying the normalized adjacent matrix $\bm{A}_{IF}$ of $\gG_{IF}$ and the embedding matrix as:
\begin{gather}
\label{eq:IFconv}
    \widetilde{\bm{A}}_{IF} = \bm{D}_{IF}^{-\frac{1}{2}} \cdot \bm{A}_{IF} \cdot \bm{D}_{IF}^{-\frac{1}{2}}, \\
    \bm{E}_{IF}^{(k+1)} = \widetilde{\bm{A}}_{IF} \cdot \bm{E}_{IF}^{(k)},
\end{gather}
where $K$ is the total GCF layers, $\bm{E}_{IF}^{(k)} \in \mathbb{R}^{(M+N) \times d}$ is embedding matrix after the $k$-th layer aggregation ($k \leq K$), and $\bm{E}_{IF}^{(0)} \in \mathbb{R}^{(M+N) \times d}$ denotes the origin I\&F graph ID embedding matrix with embedding dimension $d$. $\bm{D}_{IF}$ is the degree matrix of the adjacency matrix $\bm{A}_{IF}$.
Similarly, analogous formulas can be derived for the $\gG_{IU}$ as follows:
\begin{gather}
\label{eq:IUconv}
    \widetilde{\bm{A}}_{IU} = \bm{D}_{IU}^{-\frac{1}{2}} \cdot \bm{A}_{IU} \cdot \bm{D}_{IU}^{-\frac{1}{2}}, \\
    \bm{E}_{IU}^{(k+1)} = \widetilde{\bm{A}}_{IU} \cdot \bm{E}_{IU}^{(k)},
\end{gather}
where $\bm{E}_{IU}^{(k)} \in \mathbb{R}^{(M+N) \times d}$ is embedding matrix after the $k$-th layer aggregation, and $\bm{E}_{IU}^{(0)} \in \mathbb{R}^{(M+N) \times d}$ denotes the origin I\&U graph embedding matrix. $\bm{D}_{IU}$ is the degree matrix of $\bm{A}_{IU}$.

\subsection{Feedback-Reciprocal Contrastive Learning}
Modeling click behaviors alone is insufficient to capture diverse user intents. To mitigate the impact of I\&U, it is crucial to identify and incorporate salient information within I\&U to enrich I\&F representations. In real-world e-commerce, various user-item interactions occur, influenced by their inherent characteristics, referred to as "Impulsiveness." The goal is for the model to recognize and utilize impulsiveness to fine-tune feedback, enhancing recommendations.

Directly modeling impulsiveness from feedback results is challenging. Therefore, dividing the problem into distinct components makes it more manageable. This approach motivates the partitioning of feedback into I\&F and I\&U, based on the foundation that shared impulsiveness underlies behaviors in both categories:
\begin{equation}
    \bm{Imp} = \{imp|imp = I\&F \cap I\&U\}.
\end{equation}
Before modeling impulsiveness, we predefine six sub-relationships within the I\&F and I\&U perspectives.
Given that the User-User and Item-Item relationship graphs are absent from the dataset, we constructed $\widetilde{\bm{S}}_{IF}$ and $\widetilde{\bm{S}}_{IU}$ by using the second-order neighbors of the nodes in the user-item interaction graph from the I\&F and I\&U perspectives, respectively. This process can be formulated as:
\begin{equation}
\begin{pmatrix}
\widetilde{\bm{S}}_{IF} \\
\widetilde{\bm{S}}_{IU}\\
\end{pmatrix} =  
\begin{pmatrix}
\bm{I}+norm(\bm{A}_{IF})^2\\
\bm{I}+norm(\bm{A}_{IU})^2\\
\end{pmatrix},
\end{equation}

where $norm()$ is normalization processing. $\bm{R}_{IF}$ and $\bm{R}_{IU}$ reflect the relationships between users and items from the I\&F and I\&U perspectives, respectively. In order to make $\bm{R}_{IF}$ and $\bm{R}_{IU}$ at the same scale as $\widetilde{\bm{S}}_{IF}$ and $\widetilde{\bm{S}}_{IU}$ in the subsequent calculations, we also normalized them as:
\begin{equation}
\begin{pmatrix}
\widetilde{\bm{R}}_{IF} \\
\widetilde{\bm{R}}_{IU}\\
\end{pmatrix} =  
\begin{pmatrix}
norm(\bm{R}_{IF})\\
norm(\bm{R}_{IU})\\
\end{pmatrix}.
\end{equation}
Inspired by the adjacency matrix of the user-item graph, the normalized matrix of impulsiveness should also have user-user, item-item and user-item matrices reflecting the impulsiveness relationship, that is $\bm{Imp}_{U-U}$, $\bm{Imp}_{I-I}$, and $\bm{Imp}_{U-I}$.

As for users' impulsiveness, we argue that it consists of a user’s own social network and behavior records. Impulsiveness does not show strongly when the interests of both parties are the same, but it does show its existence when the interests of both parties conflict. In fact, people are more likely to have heated arguments with those with whom they disagree. Therefore, we define:

\begin{equation}
\begin{pmatrix}
\bm{Imp}_{U-U} \\
\bm{Imp}_{I-I}\\
\end{pmatrix} =  
\begin{pmatrix}
\widetilde{\bm{S}}_{u}^{IF}\widetilde{\bm{S}}_{u}^{IU}+\widetilde{\bm{R}}_{IF}\widetilde{\bm{R}}_{IU}^{T}\\
\widetilde{\bm{S}}_{i}^{IF}\widetilde{\bm{S}}_{i}^{IU}+\widetilde{\bm{R}}_{IF}^{T}\widetilde{\bm{R}}_{IU}\\
\end{pmatrix}.
\end{equation}

As for the impulsiveness tendency between users and items, we believe that it exists in the user's behavior records, the user's social network, and the item relationship graph.
To put it simply, it is to further explore the potential user-item relationship through the three mentioned above, and because of the existence of I\&F and I\&U from different perspectives, $\bm{Imp}_{U-I}$ has multiple forms of composition as:

\begin{equation}
\begin{pmatrix}
\bm{Imp}_{1}^{U-I} \\
\bm{Imp}_{2}^{U-I}\\
\end{pmatrix} =  
\begin{pmatrix}
\widetilde{\bm{S}}_{u}^{IF}\widetilde{\bm{R}}_{IU}+\widetilde{\bm{R}}_{IF}\widetilde{\bm{S}}_{i}^{IU}\\
\widetilde{\bm{S}}_{i}^{IF}\widetilde{\bm{R}}_{IU}^{T}+\widetilde{\bm{R}}_{IF}^{T}\widetilde{\bm{S}}_{u}^{IU}\\
\end{pmatrix}.
\end{equation}

Then, to harness the full potential of diverse structures, we offer a comprehensive formulation and a simplified expression of $\bm{Imp}$:
\begin{equation}
\bm{Imp} =  
\begin{pmatrix}
\bm{Imp}_{U-U} & \bm{Imp}_{1}^{U-I} \\
\bm{Imp}_{2}^{U-I} & \bm{Imp}_{I-I}  \\
\end{pmatrix}=\text{Joint}(\widetilde{\bm{A}}_{IF}, \widetilde{\bm{A}}_{IU})=\sum\limits_{k=0}^{2}\widetilde{\bm{A}}_{IF}^{(k)}\cdot \sum\limits_{k=0}^{2}\widetilde{\bm{A}}_{IU}^{(k)}.
\label{eq:imp}
\end{equation}

Thus, we obtain the Joint function that can extract the impulsiveness representations:

\begin{equation}
\begin{pmatrix}
\bm{E}_{IF}^{Imp} \\
\bm{E}_{IU}^{Imp}\\
\end{pmatrix} =  
\begin{pmatrix}
\text{Joint}(\widetilde{\bm{A}}_{IF}, \widetilde{\bm{A}}_{IU}) \cdot \bm{E}_{IF}^{(0)}\\
\text{Joint}(\widetilde{\bm{A}}_{IF}, \widetilde{\bm{A}}_{IU}) \cdot \bm{E}_{IU}^{(0)}\\
\end{pmatrix}.
\end{equation}

Then, we present contrastive loss between $\bm{E}_{IF}^{Imp}$ and $\bm{E}_{IU}^{Imp}$ with batch-wise contrasive loss~\cite{gutmann2010noise} as follows:
\begin{gather}
    \gL_{frcl}^{user} = \sum_{u \in \gU} -ln\frac{exp(sim(\bm{e}_{IF,u}^{Imp}, \bm{e}_{IU,u}^{Imp})/\tau)}{\sum_{u'\in\gU}exp(sim(\bm{e}_{IF,u}^{Imp}, \bm{e}_{IU,u'}^{Imp})/\tau)}, \\
    \gL_{frcl}^{item} = \sum_{i \in \gI} -ln\frac{exp(sim(\bm{e}_{IF,i}^{Imp}, \bm{e}_{IU,i}^{Imp})/\tau)}{\sum_{i'\in\gI}exp(sim(\bm{e}_{IF,i}^{Imp}, \bm{e}_{IU,i'}^{Imp})/\tau)}, \\
    \gL_{frcl} = \gL_{frcl}^{user}+\gL_{frcl}^{item},
\end{gather}
where $sim(\cdot)$ measures the similarity between two vectors, which is set as the dot product operation. $\tau$ is the temperature coefficient.

\subsection{Macro-Level Feedback Modeling}
Although feedback-reciprocal contrastive learning addresses the information exchange between the I\&F graph and the I\&U graph, it lacks the comprehensiveness of complete training data. 
To better mine information from the separated graph, we further considered feedback modeling of I\&F and I\&U graphs from a macro perspective.

GNNs' information aggregation typically occurs at the level of single micro nodes, representing a micro-level operation. 
By aggregating information from a macro standpoint, such as using clustered nodes, macro-level information can be captured. For example, modeling the user group’s interest in a certain type of item. The transformation from micro instances to macro instances can be represented as:
\begin{equation}
    \bm{Z}_{x,u} = f_{macro}(\bm{E}_{x,u}^{(0)}), \quad \bm{Z}_{x,i} = f_{macro}(\bm{E}_{x,i}^{(0)}),
\end{equation}
\begin{equation}
    \bm{C}_{x}= [\bm{Z}_{x,u}, \bm{Z}_{x,i}],
\end{equation}
where macro transformation function $f_{macro}(\cdot)$ is adopted with the K-means clustering algorithm~\cite{hamerly2003learning} in this paper, $x$ represents either I\&F or I\&U, $\bm{Z}_{x,u}$ and $\bm{Z}_{x,i}$ are the clustering centroid representations, $\bm{C}_{x}\in \mathbb{R}^{Q\times d}$ denotes the macro centroid representation matrix, and $Q$ denotes the total number of centroids. Then, the challenge is to aggregate macro-level information to nodes in a way that produces results similar to those from GNNs~\cite{kipf2016semi,chen2024macro}. Intuitively, we would adjust the GNN propagation mechanism for macro-level aggregation. However, understanding the intricacies of GNN information propagation and using it to guide macro-level aggregation is difficult. 
Therefore, we turn to propose leveraging the differences between results from various GNN layers and averaging them to infer the direction of GNN information propagation, denoted as:
\begin{equation}
    \bm{V}_{x} = \frac{1}{K}\sum\limits_{i=0}^{K-1}(\bm{E}_{x}^{(i+1)}-\bm{E}_{x}^{(i)}),
\end{equation}
where $K$ is the number of GNN layers. 
In GNNs, the direction of node updates inferred from the graph structure (denoted as $\bm{V}_{x}$) provides valuable insight into the decision-making process. If macro-level information aggregation follows these directional cues, it validates the alignment between macro and micro outcomes. To achieve this, we calculate the similarity between the update directions of individual nodes and cluster centroids. This determines the weights for each cluster centroid during macro-level aggregation for each node. For conciseness, we denote the computational outcome of $\bm{V}_{x}\bm{C}_{x}^{T}$ as $\bm{H}_{x}$. Subsequently, we employ the derived weights $\bm{W}_{x}$ to aggregate macro-level information across all nodes within the graph:
\begin{equation}
    \bm{W}_{x} = \frac{\bm{H}_{x}(i,j)}{\sqrt{\sum\limits_{j=1}^{Q}\bm{H}_{x}(i,j)^{2}}},\quad \bm{E}^{macro}_{x} = \bm{E}_{x}^{(0)}+\frac{1}{Q}\bm{W}_{x}\bm{C}_{x}.
\end{equation}

Upon acquiring the $\bm{E}_{x}^{macro}$, we employ the aggregated outcomes from nodes associated with two different perspectives as positive samples and the remaining ones deemed as negative samples, calculated with the widely adopted InfoNCE loss~\cite{chen2020simple,you2020graph}. This formulation yields the ensuing equation:
\begin{equation}
    \mathcal{L}_{macro}^{x} = \text{InfoNCE}(\bm{E}_{x,u}^{(K)},\bm{E}^{macro}_{x,u})+\mu\cdot\text{InfoNCE}(\bm{E}_{x,i}^{(K)},\bm{E}^{macro}_{x,i}),
\end{equation}
\begin{equation}
    \mathcal{L}_{macro} = \mathcal{L}_{macro}^{IF}+\mathcal{L}_{macro}^{IU},
\end{equation}
where $\mu$ is a hyper-parameter to balance the weight of loss functions between the user and item sides.

\subsection{Optimization and Inference}
\subsubsection{Distance Regularization}
In the allocation of distinct representations to various graph structures, the extent of decoupling between these representations becomes a focal point of concern. To optimize this decoupling, we adopt an indirect strategy aimed at regularizing these representations, thus augmenting their expressiveness. This objective is realized through the introduction of disparity into the probability distributions governing the update directions of the respective GNNs. A prudent design approach entails the utilization of Jensen-Shannon divergence~\cite{lin1991divergence} (short as JSD) for the explicit purpose of regularization:
\begin{equation}
%\mathcal{L}_{far}=\frac{1}{2}KL(D_{IF}||\frac{D_{IF}+D_{IU}}{2})+\frac{1}{2}KL(D_{IU}||\frac{D_{IF}+D_{IU}}{2})
\mathcal{L}_{dis} = -JSD(\bm{V}_{IF}, \bm{V}_{IU}).
\end{equation}

\subsubsection{Overall Objective Function}
We apply the pairwise BPR loss~\cite{rendle2009bpr} to train both the I\&F GCF and the I\&U GCF. Then after including the I\&F recommendation loss, I\&U recommendation loss, feedback-reciprocal contrastive loss, macro-level feedback modeling loss, and distance regularization, we have:
\begin{equation}
    \mathcal{L}_{bpr}^{IF} = -\sum\limits_{(u,i,j)\in\mathcal{O}}ln\sigma(\hat{y}_{ui}^{IF}-\hat{y}_{uj}^{IF}),
\end{equation}

\begin{equation}
    \mathcal{L}_{bpr}^{IU} = -\sum\limits_{(u,i,j)\in\mathcal{O}}ln\sigma(\hat{y}_{ui}^{IU}-\hat{y}_{uj}^{IU}),
\end{equation}

\begin{equation}
    \gL = \mathcal{L}_{bpr}^{IF} + \mathcal{L}_{bpr}^{IU}+ \lambda_1 \cdot \mathcal{L}_{frcl}+\lambda_2\cdot \mathcal{L}_{macro}+\lambda_3\cdot \mathcal{L}_{dis},
\end{equation}
where the parameters $\lambda_1$, $\lambda_2$, and $\lambda_3$ as hyperparameters aimed at regulating the relative impacts of feedback-reciprocal contrastive learning, macro-level feedback modeling, and distance regularization, respectively.

\subsubsection{Inference}
Upon completion of the \model training, we acquire the $\mE_{IF}$ and $\mE_{IU}$. During the inference stage, when considering a user $u$ and an item $i$, the recommendation outcomes $\hat{y}_{ui}$ can be generated by the model of the I\&F GCF side by calculating the similarity between $u$ and $i$ and ranking accordingly.

%%%%%%%%%%%%%%%%%%
\section{Experiments}
In this section, we conduct comprehensive experiments to demonstrate the effectiveness of the proposed \model method. Specifically, we aim to answer the following research questions:
%\begin{itemize}[leftmargin=*]
\textbf{RQ1:} How does \model perform compared with state-of-the-art interaction recommendation models?
\textbf{RQ2:} Whether \model can recommend more fascinated items and less unfascinated items?
\textbf{RQ3:} If each tailored component can positively impact the recommendation performance?
\textbf{RQ4:} How does \model perform in the real-world industrial recommender system?
%\end{itemize}

\subsection{Experimental Setup}
\subsubsection{Datasets}
We evaluate our method on four benchmark datasets: \textbf{MovieLens}~\cite{harper2015movielens}, \textbf{Douban}~\cite{wu2017sequential}, \textbf{Beauty}~\cite{mcauley2015image}, \textbf{KuaiRec}~\cite{gao2022kuairec}, and a billion-scale \textbf{industrial} dataset with 0.17 billion users and 0.31 billion items collected from Taobao platform. 
The statistics of these datasets are shown in Table \ref{tab:stats}. 
We follow the commonly adopted settings in previous works~\cite{wang2019neural,he2020lightgcn} to randomly split the datasets into training and testing sets. 
%\textbf{Dataset setup}. 
To partition the interaction graph, for KuaiRec, 
we consider interactions with a completion rate greater than 50\% as I\&F and those less than 50\% as I\&U
feedback. For the other public benchmark datasets, we regard the interactions with ratings greater than or equal to 4 as I\&F feedback and those less than 4 as I\&U feedback~\cite{zhou2018deep}. For the industrial dataset, we view the interactions with page stay time more than 4 seconds as I\&F feedback and those less or equal to 4 seconds as I\&U feedback.

\begin{table}[htbp]
  \centering
  \small
  \caption{Statistics of the experimental datasets.}
  \resizebox{\linewidth}{!}{
  \setlength{\tabcolsep}{2mm}{
    \begin{tabular}{c|c|c|c|c}
    \toprule
    Dataset & \# Users & \# Items & \# Interactions &  \# Negative Type \\
    \midrule
    MovieLens & 6,040 & 3,952 & 1,000,209 & Low rating \\
    Douban & 2,848 & 39,586 & 894,886 & Low rating \\
    Beauty & 324,038 & 32,586 & 361,605 & Low rating \\
    KuaiRec & 7,176 & 10,728 & 12,530,806 & Short completion\\
    \midrule
    Industrial & $\sim$0.17 Billion & $\sim$0.31 Billion & $\sim$118 Billion & Quick glance \\
    \bottomrule
    \end{tabular}%
  }
}
  \label{tab:stats}%
\end{table}%

\begin{table*}[t]
  \centering
  \caption{Overall performance on four experimental datasets. ``(I\&F)'' denotes the model trained only on the I\&F graph. The best and second-best results are highlighted in \textbf{bold} font and \underline{underlined}.}
  %\vspace{-0.5em}
  \resizebox{0.85\linewidth}{!}{
    \begin{tabular}{c|cc|cc|cc|cc|cc}
    \toprule
    %\toprule
    \multirow{2}[2]{*}{Method} & \multicolumn{2}{c|}{MovieLens} & \multicolumn{2}{c|}{Douban} & \multicolumn{2}{c|}{Beauty} & \multicolumn{2}{c}{KuaiRec} & \multicolumn{2}{c}{Industrial} \\
          & Recall & NDCG  & Recall & NDCG  & Recall & NDCG  & Recall & NDCG & Recall & NDCG \\
    \midrule
    BUIR  & 0.1672 & 0.2978 & 0.0698 & 0.2014 & \underline{0.1107} & 0.0788 & 0.0384 & 0.3489 & 0.1033 & 0.2034 \\
    DirectAU & 0.1086 & 0.1375 & 0.0189 & 0.0377 & 0.0980 & 0.0669 & 0.0158 & 0.1107 & 0.0975 & 0.1389 \\
    LightGCN & 0.2251 & 0.3636 & 0.0755 & 0.2110 & 0.0996 & 0.0786 & \underline{0.0726} & \underline{0.6131} & 0.1542 & 0.2763 \\
    LightGCN-(I\&F) & 0.2074 & 0.3513 & 0.0737 & 0.2183 & 0.1023 & 0.0763 & 0.0725 & 0.6129 & 0.1343 & 0.2564 \\
    %\midrule
    SGL   & 0.2377 & \underline{0.3839} & 0.0855 & 0.2136 & 0.0987 & 0.0761 & 0.0530 & 0.4614 & 0.1693 & 0.2678 \\
    SGL-(I\&F) & 0.2259 & 0.3709 & 0.0899 & \underline{0.2396} & 0.0931 & 0.0741 & 0.0530 & 0.4546 & 0.1411 & 0.2319 \\
    SimGCL & \underline{0.2421} & 0.3750 & \underline{0.0911} & 0.2279 & 0.0912 & 0.0723 & 0.0545 & 0.4888 & 0.1789 & \underline{0.2811} \\
    SimGCL-(I\&F) & 0.2208 & 0.3505 & 0.0809 & 0.2115 & 0.0834 & 0.0639 & 0.0543 & 0.4882 & 0.1476 & 0.2601 \\
    NCL   & 0.2331 & 0.3787 & 0.0755 & 0.2206 & 0.1048 & \underline{0.0809} & 0.0714 & 0.6019 & \underline{0.1801} & 0.2756 \\
    NCL-(I\&F) & 0.1816 & 0.3116 & 0.0756 & 0.2095 & 0.1025 & 0.0759 & 0.0710 & 0.6028 & 0.1489 & 0.2543 \\
    SiReN & 0.2075 & 0.3520 & 0.0699 & 0.2089 & 0.1066 & 0.0788 & 0.0701 & 0.5948 & 0.1765 & 0.2793 \\
    \midrule
    \textbf{\model} (ours) & \textbf{0.2533} & \textbf{0.4027} & \textbf{0.0969} & \textbf{0.2586} & \textbf{0.1198} & \textbf{0.0864} & \textbf{0.0767} & \textbf{0.6404} & \textbf{0.1919} & \textbf{0.3031} \\
    Improvement (\%) & 4.62\% & 4.89\% & 6.36\% & 7.92\% & 8.22\% & 6.79\% & 5.64\% & 4.45\% & 6.55\% & 7.82\% \\
    \bottomrule
    %\bottomrule
    \end{tabular}%
    }
    %\vspace{-0.5em}
  \label{tab:main_tab}%
\end{table*}%

\subsubsection{Baselines}
We compare the proposed \model with seven representative state-of-the-art graph collaborative filtering recommendation models as follows: (i) Representative GCF models: \textbf{BUIR}~\cite{lee2021bootstrapping}, \textbf{DirectAU}~\cite{wang2022towards}, and \textbf{LightGCN}~\cite{he2020lightgcn}. (ii) SSL-enhanced GCF models: \textbf{SGL}~\cite{wu2021self}, \textbf{SimGCL}~\cite{yu2022graph}, and \textbf{NCL}~\cite{lin2022improving}. (iii) User feedback-based GCF model:  \textbf{SiReN}~\cite{seo2022siren}.

\subsubsection{Parameter Setting \& Evaluation Metrics}
The embedding size is fixed to 64 for all models~\cite{he2020lightgcn}, and the embedding parameters are initialized with the Xavier method~\cite{glorot2010understanding}.
We adopt the Adam optimizer~\cite{kingma2014adam}  and a default batch size of 2048. We choose the widely used all-rank Recall@\textit{N} and NDCG@\textit{N} (\textit{N}=20) as the evaluation metrics~\cite{wang2019neural,he2020lightgcn}. For fair comparisons, we run the experiments \textit{five} times with different random seeds and report the average results to prevent extreme cases.

\subsection{Offline Evaluation (RQ1)}
To verify the superiority of \model, we conduct the offline experiment on the interaction recommendation performance of \model, baseline methods, and their variants that are directly trained ignoring the I\&U graph.
Table \ref{tab:main_tab} illustrates the offline evaluation results.
From these results, we have the following observations:

%\begin{itemize}[leftmargin=*]
    \textbf{\model outperforms all state-of-the-art methods on all datasets.} Specifically, Recall and NDCG metrics are improved over 6.28\% and 6.37\% for the experimental datasets on average, respectively.
    Note that \model gains 6.55\% and 7.82\% on Recall and NDCG respectively of the billion-scale industrial dataset. These observations indicate that comprehensively considering fascinated interactions and unfascinated interactions can better model users' interaction intent. Besides, incorporating the feedback-partitioned graph collaborative filtering, feedback-reciprocal contrastive learning, and macro-level feedback modeling with \model can address the highlighted dilemma and performance decays caused by splitting the complete graph.
    
    \textbf{The consideration of I\&U graph is helpful to the model performance.} Compared with training on the whole interacted graph, LightGCN, SGL, SimGCL, and NCL only trained on the I\&F graph are more likely to lead to a significant decline in model performance. These results verify that simply removing all the I\&U interactions for model training will result in sub-optimal performance. Thus, an effective solution for modeling user feedback through the I\&U graph and the I\&F graph simultaneously is needed, which verifies the necessity of \model's feedback-reciprocal contrastive learning and macro-level feedback modeling.
%\end{itemize}

\begin{table}[tbp]
  \centering
  \caption{Fascinated \& Unfascinated recommendation comparison with top-3 performed baselines ($\uparrow$: the higher, the better; $\downarrow$: the lower, the better).}
  %\vspace{-0.5em}
  \resizebox{0.925\linewidth}{!}{
    \begin{tabular}{c|cc|cc}
    \toprule
    \multirow{2}[2]{*}{Model} & \multicolumn{2}{c|}{Fascinated Rec. ($\uparrow$)} & \multicolumn{2}{c}{Unfascinated Rec. ($\downarrow$)} \\
          & Recall@20 & NDCG@20 & Recall@20 & NDCG@20 \\
    \midrule
    \model & \textbf{0.1142} & \textbf{0.2294} & \textbf{0.0473}  & \textbf{0.0527}  \\
    \midrule
    SGL   & \underline{0.1092}  & \underline{0.2062}  & 0.0607  & 0.0752  \\
    SimGCL & 0.1023  & 0.1966  & 0.0501  & \underline{0.0559}  \\
    NCL   & 0.0885  & 0.1855  & \underline{0.0478}  & 0.0574  \\
    \bottomrule
    \end{tabular}%
    }
    %\vspace{-1em}
  \label{tab:case_study}%
\end{table}%

\subsection{Recommendation Preference (RQ2)}
We further conduct the case study on Douban to observe the recommendation performance of \model on fascinated and unfascinated items. The recommendation comparison results of the user-fascinated interactions and the user-unfascinated interactions are presented in Table~\ref{tab:case_study}. 

From these results, we can find that \textbf{\model can accurately recommend more user-fascinated items while effectively reducing the recommendation of user-unfascinated items}.
Specifically, compared with the top-performed baselines (SGL, SimGCL, and NCL), \model can improve the overall recommendation performance by recalling more fascinating items. Then, for the unfascinated recommendation, \model recalls less unfascinating items and diminishes their influence in the ultimate recommendation outcomes, indicating that with the help of feedback-reciprocal contrastive learning and macro-level feedback modeling, \model can effectively model the unfascinated interactions while reducing the recommendation of them.

\subsection{Ablation Study (RQ3)}
In order to verify the effectiveness of the designed components of \model, we conduct the ablation study on \model by comparing with its three variants: \model without feedback-reciprocal contrastive learning~(\model-\textit{w/o frcl}), \model without macro-level feedback modeling~(\model-\textit{w/o macro}) and \model without distance regularization~(\model-\textit{w/o dis}).

\begin{figure}[tbp]
    \centering
    \includegraphics[width=\linewidth, trim=0cm 0cm 0cm 0cm,clip]{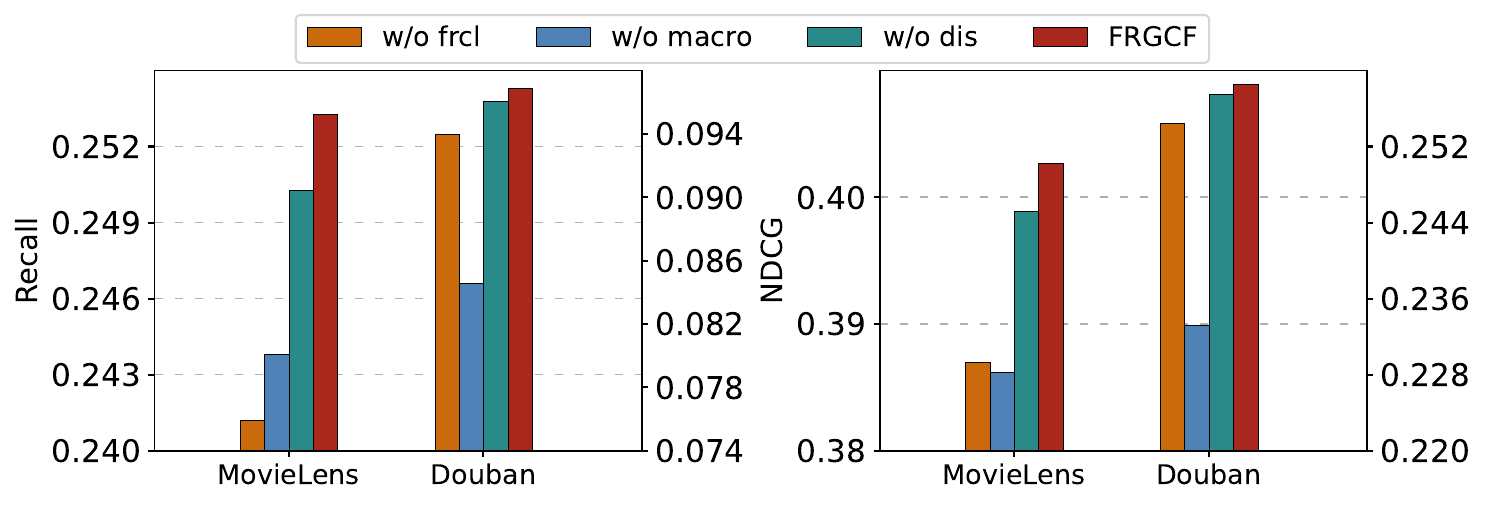}
    \caption{Ablation study results on three \model variants.}
    \label{fig:ablation}
\end{figure}

Upon analyzing the results presented in Figure~\ref{fig:ablation}, it can be inferred that when compared to the \model-\textit{w/o frcl} and \textit{w/o macro} variant, it is evident that \model yields significant improvement, thereby substantiating the need for feedback-reciprocal information exchange between the I\&F graph and I\&U graph and the support from macro-level to micro-level information. Furthermore, compared with \model-\textit{w/o dis}, the enhancement derived from distance regularization substantiates the imperative of representation disentanglement between the I\&F graph and the I\&U graph.

\subsection{Online Evaluation (RQ4)}
In order to verify the capability of \model in real system application, we further conducted an online A/B test for \model on Taobao's industrial recommender system, where \model serves as a recall model, replacing the currently online recalling baseline model. Table~\ref{tab:abtest} displays the average relative performance gain over four consecutive weeks for twenty million platform users.

From the results, we can have the following observations: Firstly, compared to the baseline model, \model demonstrates a performance improvement of 1.35\% for PCTR and 0.91\% for UCTR, suggesting that users are more willing to click since \model recommends items that truly fascinate them.
Furthermore, GMV and Stay Time increase by 1.61\% and 2.89\%, respectively, indicating that \model encourages users to engage more deeply with the recommended items, such as making purchases or exploring item details and videos.
Lastly, \model needs only slightly additional response time cost as the baseline model, demonstrating that \model significantly improves performance while maintaining efficiency.

\begin{table}[tbp]
  \centering
  \caption{Results of online A/B tests in the industrial platform.}
  %\vspace{-0.5em}
  \resizebox{\linewidth}{!}{
    \begin{tabular}{l|ccccc}
    \toprule
    A/B Test & PCTR & UCTR & GMV & StayTime & ResTime\\
    \midrule
    v.s. Baseline & +1.35\% & +0.91\% & +1.61\% & +2.89\% & +0.19\%\\
    \bottomrule
    \end{tabular}%
    %\vspace{-0.5em}
   }
  \label{tab:abtest}%
\end{table}%

%%%%%%%%%%%%%%%%%%%%
\section{Conclusion}
In this paper, we first highlight the seesaw dilemma in graph collaborative filtering models of considering only the I\&F interactions or both the I\&F and I\&U interactions. To address this issue, we proposed \model, the Feedback Reciprocal Graph Collaborative Filtering model, which introduced novel feedback-partitioned graph collaborative
filtering, feedback-reciprocal contrastive learning, and macro-level feedback modeling architectures for user and item representation learning to recommend more user-fascinated items and fewer user-unfascinated items. Extensive offline and online experiments verify the effectiveness of \model, which demonstrated significant recommendation improvement compared to state-of-the-art methods and the applicability of \model in industrial utilization.

\begin{acks}
This work was supported in part by the National Natural Science Foundation of China (Grant No. 62272200, U22A2095, 61932010).
\end{acks}

%\clearpage
\bibliographystyle{ACM-Reference-Format}
\bibliography{reference.bib}

\end{document}